\documentclass[conference]{IEEEtran}
\IEEEoverridecommandlockouts

\usepackage{cite}
\usepackage{amsmath,amssymb,amsfonts}
\usepackage{algorithmic}
\usepackage{graphicx}
\usepackage{times}
\usepackage{courier}
\usepackage{bm}
\usepackage{times}
\usepackage{url}
\usepackage[small]{caption}
\usepackage{graphicx}
\usepackage{booktabs}
\usepackage{amsmath}
\usepackage{amsthm}
\usepackage{algorithm}
\usepackage{algorithmic}
\usepackage[switch]{lineno}
\def\BibTeX{{\rm B\kern-.05em{\sc i\kern-.025em b}\kern-.08em
    T\kern-.1667em\lower.7ex\hbox{E}\kern-.125emX}}
\begin{document}

\title{Report on Methods and Applications for Crafting 3D Humans
}

\author{\IEEEauthorblockN{Lei Liu, Ke Zhao, Bournemouth University\\}
}

\maketitle

\begin{abstract}
This paper presents an in-depth exploration of 3D human model and avatar generation technology, propelled by the rapid advancements in large-scale models and artificial intelligence. The paper reviews the comprehensive process of 3D human model generation, from scanning to rendering, and highlights the pivotal role these models play in entertainment, VR, AR, healthcare, and education. We underscore the significance of diffusion models in generating high-fidelity images and videos. It emphasizes the indispensable nature of 3D human models in enhancing user experiences and functionalities across various fields. Furthermore, this paper anticipates the potential of integrating large-scale models with deep learning to revolutionize 3D content generation, offering insights into the future prospects of the technology. It concludes by emphasizing the importance of continuous innovation in the field, suggesting that ongoing advancements will significantly expand the capabilities and applications of 3D human models and avatars.
\end{abstract}


\section{Introduction}

The rapid advancement of large-scale models~\cite{rombach2022high} and artificial intelligence has significantly transformed the landscape of digital content creation. Among the various innovations, 3D content generation technology stands out due to its extensive applications across numerous domains. This paper focuses on the generation of 3D human models and avatars, exploring their technical foundations, applications, challenges, and future prospects.
Diffusion models have shown strong capabilities in high-fidelity images and video generation~\cite{li2023layerdiffusion,hertz2022prompt,ruiz2023dreambooth,ho2022imagenvideo,brooks2023instructpix2pix,couairon2022diffedit,ramesh2021zero}. Applying pre-trained diffusion models to 3D content generation can significantly reduce the demand for computational resources and the reliance on 3D datasets. 3D generation~\cite{tang2023dreamgaussian,li2024art3d,lin2023magic3d,xiong2023get3dhuman,shi2023zero123++,raj2023dreambooth3d,sun2023dreamcraft3d,li2023archi,li2023efficient,ranftl2020towards} involves creating three-dimensional digital representations of objects, beings, or environments through sophisticated algorithms and tools. This technology has gained prominence in entertainment, virtual reality (VR), augmented reality (AR), healthcare, and education, offering new dimensions of interactivity and realism. Specifically, 3D human models and avatars play pivotal roles in enhancing user experience and functionality in these fields.

The process of generating 3D human models~\cite{zhang2023dreamface,zhang2023dreamface,jiang2023avatarcraft,weng2022humannerf,xiong2023get3dhuman} typically includes scanning, modeling, and rendering. Scanning technologies capture the geometric and textural details of a human subject using 3D scanners or multi-view camera systems. The captured data is then processed through modeling techniques such as mesh modeling and voxel modeling to create a high-fidelity 3D representation. Finally, rendering techniques, including ray tracing and global illumination, convert these models into visually realistic images. These technologies collectively enable the creation of lifelike digital humans that can be animated and manipulated in various digital environments.

In entertainment and media, 3D human models are indispensable. Films and video games leverage these models to bring characters to life, enhancing storytelling and visual impact. For instance, blockbuster movies like "Avatar" and "The Avengers" extensively use 3D modeling to create immersive visual effects and realistic characters. Similarly, in video games, these models contribute to creating engaging and interactive experiences, allowing players to immerse themselves in virtual worlds.

Virtual reality (VR) and augmented reality (AR) applications also benefit greatly from 3D human models. In VR, users can interact with highly detailed and realistic human avatars, making the virtual experience more immersive and engaging. AR applications overlay 3D human models onto the real world, providing enhanced interactivity for educational, training, and entertainment purposes. The realism and interactivity provided by 3D human models are crucial for the effectiveness and appeal of VR and AR experiences.

Avatars, or digital representations of users, are another significant application of 3D content generation technology. Avatars are extensively used in social media, online communication, and virtual environments. Creating avatars involves face recognition and modeling, skeletal animation, and expression capture. These avatars can be customized to reflect the user's appearance, personality, and preferences, providing a personalized and engaging digital presence. Social media platforms like Meta's Horizon Worlds and Snapchat's Bitmoji allow users to create and interact through avatars, enhancing social interactions and personal expression. In online education and training, avatars represent teachers and students, facilitating interactive and immersive learning experiences. Virtual meetings and conferences also employ avatars to create more engaging and lifelike interactions, improving communication and collaboration among remote participants.

The rapid evolution of artificial intelligence and machine learning has ushered in a transformative era for 3D modeling, with text-to-image prior-guided text-to-3D modeling at the forefront of this technological revolution. This innovative approach leverages the power of natural language processing to intelligently generate 3D models that correspond to textual descriptions, significantly expanding the horizons and possibilities of 3D content creation. Within this domain, a variety of more nuanced and specialized applications have emerged, including personalized avatar creation.

The application of text-to-avatar allows users to craft unique 3D avatars through simple textual input, reflecting not only their physical characteristics but also their personality and emotional states. Across social media platforms, gaming, and virtual reality environments, this technology offers a novel mode of self-expression, enhancing user interaction and immersive experiences.

This paper delves into text-guided 3D model generation models that are predicated on text-to-image priors, examining their underlying principles, technical frameworks, and practical effectiveness across various application scenarios. Through these investigations, we aim to provide new perspectives and insights into the field of 3D modeling, propelling continuous technological innovation and advancement.

Looking ahead, the integration of large-scale models and deep learning into 3D content generation holds great promise. AI can enhance the accuracy and efficiency of 3D modeling processes, enabling the creation of more realistic and complex models with less manual intervention. Deep learning techniques can improve facial recognition, expression capture, and animation, making avatars more lifelike and expressive.

\section{3D Technologies}

\subsection{3D Representation}

3D representation is vital in computer graphics, enabling the digital visualization and manipulation of three-dimensional objects and scenes. Different methods of 3D representation offer varying degrees of detail, efficiency, and application suitability. Here, we delve into several key types of 3D representations, including Explicit Representations, Point Clouds, and Voxels.

\paragraph{Explicit Representations}

Explicit representations define 3D objects with precise mathematical descriptions. This category includes mesh-based methods, which represent surfaces using vertices, edges, and faces. Meshes are widely used due to their flexibility and efficiency in rendering.

\paragraph{Meshes, Point Clouds and Voxels}
A mesh is composed of polygons, typically triangles, which are connected by their edges and vertices to form the surface of a 3D object. Meshes are favored in applications requiring detailed and smooth surfaces, such as video games, animations, and simulations. They allow for efficient rendering~\cite{botsch2010polygon, shirman1987local} and easy manipulation but can become complex when representing highly detailed surfaces.

Point clouds~\cite{aliev2020neural, kopanas2021point, kopanas2021point} represent 3D shapes as collections of discrete points in space. Each point has specific (x, y, z) coordinates and often includes attributes like color or normal vectors. Point clouds are useful in scanning and reconstruction applications, as they capture surface geometry directly from real-world scans. They are employed in fields like autonomous driving, where LiDAR sensors generate point cloud data to represent the environment. However, point clouds can be sparse and may require significant processing to convert into other forms of representation, like meshes or voxel grids, for further use. The lack of explicit connectivity between points can complicate operations like surface reconstruction and rendering.

Voxels~\cite{liu2020neural, sun2022direct, maturana2015voxnet}, or volumetric pixels, are the 3D equivalent of pixels in a 2D image. They divide the 3D space into a grid of small cubes, each storing information about the material present at that location. Voxel grids are suitable for applications requiring uniform space representation. They handle complex topologies and internal structures of objects well. However, their memory consumption is high, as high-resolution voxel grids can become very large. This makes them less efficient for detailed surface representations compared to meshes. Advancements in sparse voxel representations and efficient storage techniques are mitigating these issues.

\subsection{NeRF and 3DGS}

Neural Radiance Fields (NeRF)~\cite{mildenhall2020nerf, niemeyer2021giraffe, barron2022mip, chen2021mvsnerf} synthesize novel views of complex 3D scenes from sparse sets of 2D images. NeRF represents a scene using a continuous 5D function mapping spatial coordinates and viewing directions to color and density values. This enables photorealistic image creation from new viewpoints by interpolating between input images. NeRF employs a fully-connected neural network to optimize a volumetric scene function, accumulating colors and densities to form a 2D image. The process involves marching rays through the scene, sampling 3D points, and predicting color and density values using the neural network, allowing high-fidelity reconstructions with intricate details and realistic lighting. Classic NeRF has limitations, such as long training times and the need to train a new model for each scene. Extensions like InstantNeRF~\cite{muller2022instant}, which uses a multiresolution hash table to accelerate training, and PixelNeRF~\cite{yu2021pixelnerf}, which leverages convolutional neural networks for view synthesis from a single image, address these challenges.

3D Gaussian Splatting~\cite{abdal2023gaussian, chen2023text, tang2023dreamgaussian, liu2023humangaussian, yi2023gaussiandreamer} is a sophisticated volumetric rendering technique used for creating and manipulating intricate visual effects within a three-dimensional space. This method involves "splatting" Gaussian distributions throughout the volume data to simulate the impact of light sources, material properties, and geometric shapes on a scene. Each Gaussian distribution, or "splat," represents a minuscule volumetric element of the scene, encapsulating attributes such as color, density, and transparency. By blending these splats, 3D Gaussian Splatting can produce highly realistic imagery, making it a valuable tool for applications in computer graphics, visual effects, and scientific visualization. Potential applications range from virtual property tours and urban planning to creating photorealistic avatars for telepresence in VR environments.

\subsection{Diffusion Models}

Diffusion models are generative models based on the idea of reversing a diffusion process, transforming data from a simple initial distribution to a complex target distribution. The diffusion process involves gradually adding noise to data, while the reverse process denoises it to generate new samples.

\paragraph{Forward Process (Diffusion)}

The forward process can be described as a Markov chain where noise is added to the data at each time step. This is represented by the following equation:

\[
q(x_t | x_{t-1}) = \mathcal{N}(x_t; \sqrt{1 - \beta_t} x_{t-1}, \beta_t \mathbf{I})
\]

Here:
- \( x_t \) is the data at time step \( t \).
- \( \beta_t \) is a variance schedule controlling the amount of noise added.
- \( \mathcal{N} \) denotes a Gaussian distribution.

The forward process starts from the original data \( x_0 \) and progressively adds noise:

\[
q(x_t | x_0) = \mathcal{N}(x_t; \sqrt{\bar{\alpha}_t} x_0, (1 - \bar{\alpha}_t) \mathbf{I})
\]

where:
- \( \bar{\alpha}_t = \prod_{s=1}^t (1 - \beta_s) \).

\paragraph{Reverse Process (Denoising)}

The reverse process learns to denoise the data, moving from the noisy data \( x_t \) back to the original data \( x_0 \). The reverse process is parameterized by a neural network \( p_\theta \):

\[
p_\theta(x_{t-1} | x_t) = \mathcal{N}(x_{t-1}; \mu_\theta(x_t, t), \sigma^2_t \mathbf{I})
\]

Here, \( \mu_\theta(x_t, t) \) is the mean predicted by the neural network, and \( \sigma^2_t \) is typically fixed.

\paragraph{Training Objective}

The training objective is to minimize the difference between the true reverse process and the model's predictions. This can be formulated as:

\[
L_\text{simple} = \mathbb{E}_{t, x_0, \epsilon} \left[ \left\| \epsilon - \epsilon_\theta(x_t, t) \right\|^2 \right]
\]

where:
- \( \epsilon \) is the noise added in the forward process.
- \( \epsilon_\theta \) is the neural network's prediction of the noise.

\paragraph{Sampling}

To generate new samples, the model starts from random noise \( x_T \) and iteratively applies the reverse process:

\[
x_{t-1} = \frac{1}{\sqrt{\alpha_t}} \left( x_t - \frac{\beta_t}{\sqrt{1 - \bar{\alpha}_t}} \epsilon_\theta(x_t, t) \right) + \sigma_t z
\]

where \( z \sim \mathcal{N}(0, \mathbf{I}) \) is Gaussian noise.

In summary, diffusion models leverage a forward process that adds noise to the data and a learned reverse process that removes noise, enabling the generation of complex data distributions from simple initial noise.

\begin{figure}[t]
	\centering

	\includegraphics[width=\linewidth]{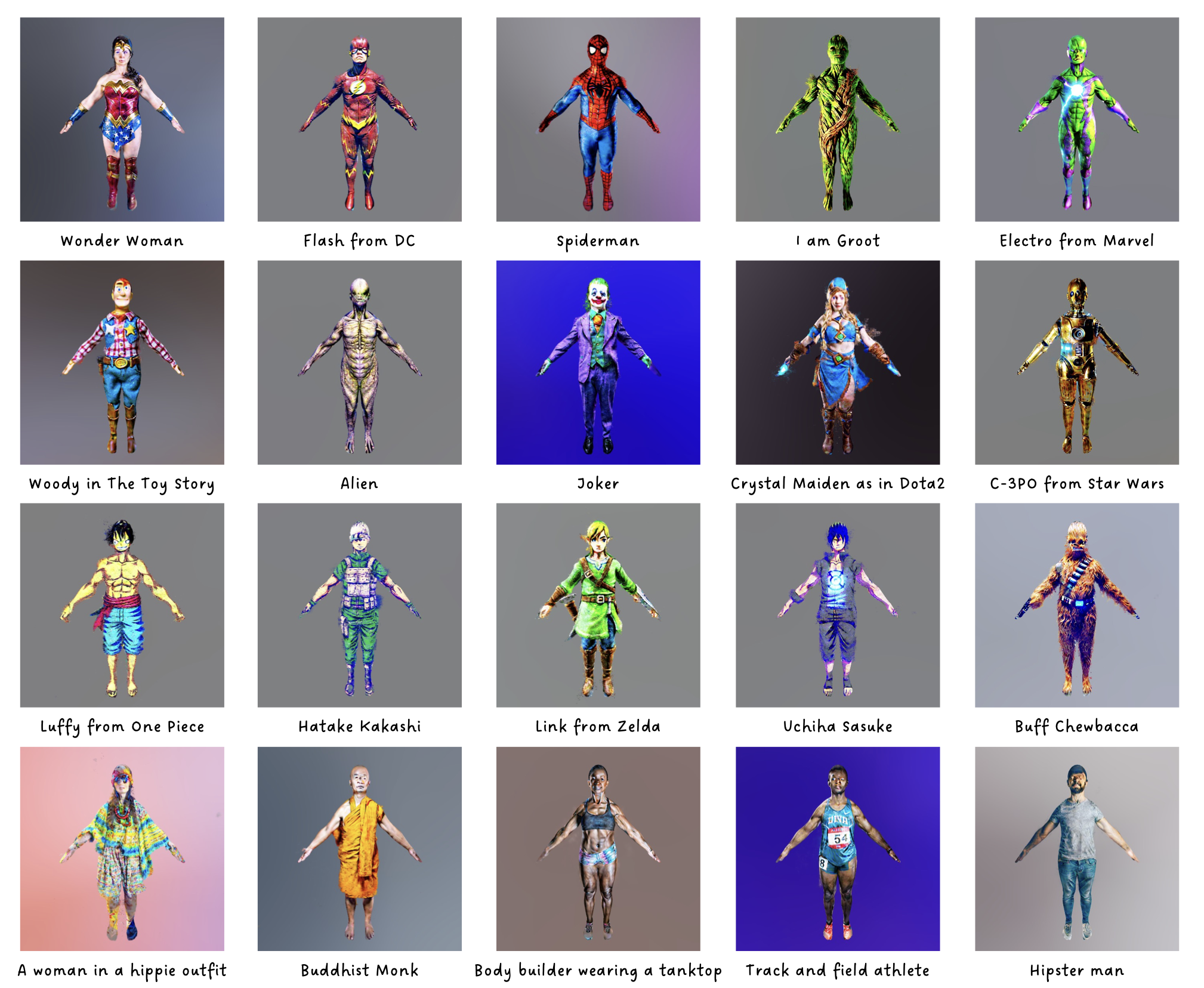}

	\caption{DreamAvatar~\cite{cao2023dreamavatar} can generate high-quality geometry and texture for any type of human avatar.}
	\label{fig:01}
\end{figure}

\begin{figure}[t]
	\centering

	\includegraphics[width=\linewidth]{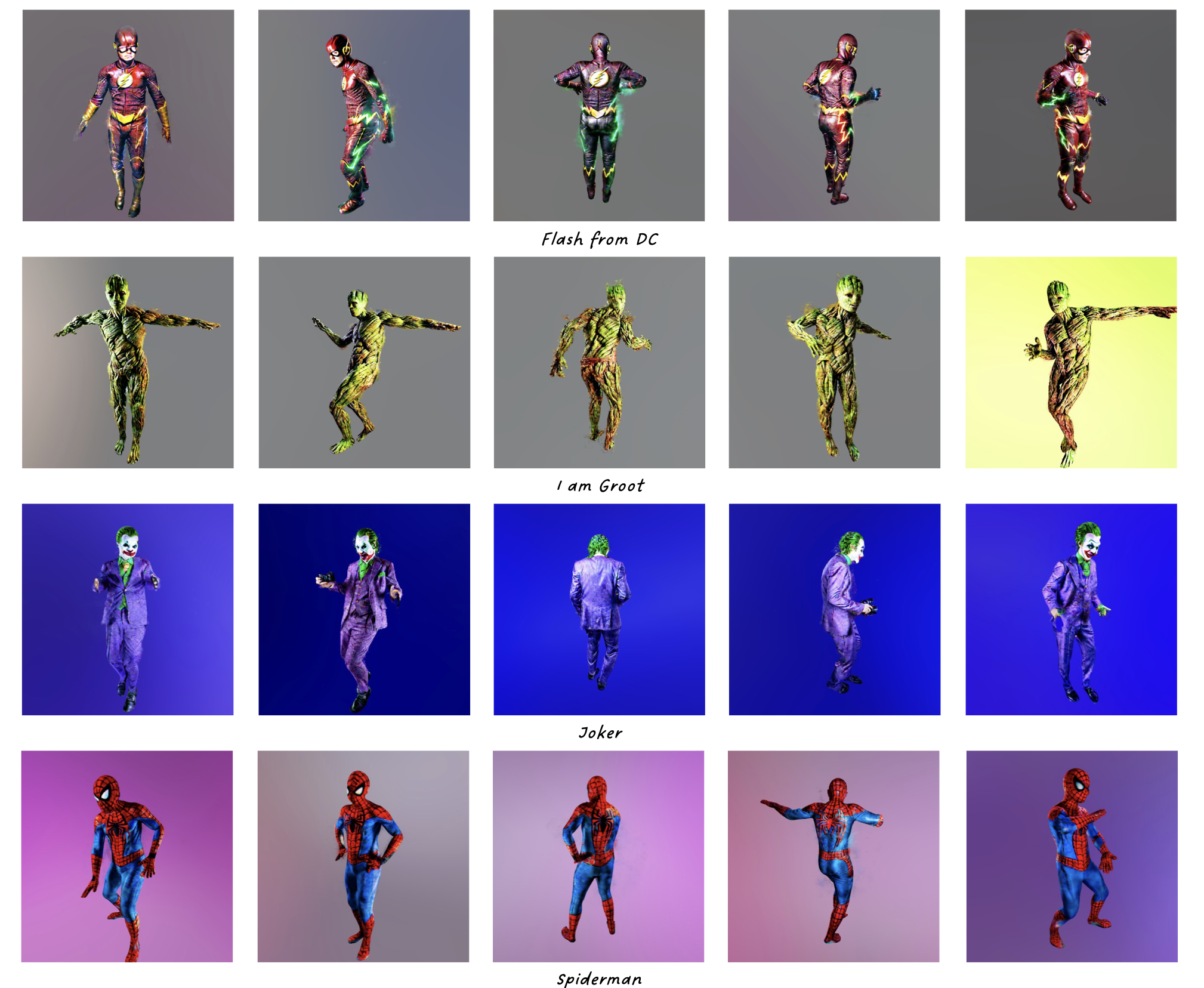}

	\caption{DreamAvatar can handle and control the 3D generation with any pose.}
	\label{fig:02}
\end{figure}

\begin{figure}[t]
	\centering

	\includegraphics[width=\linewidth]{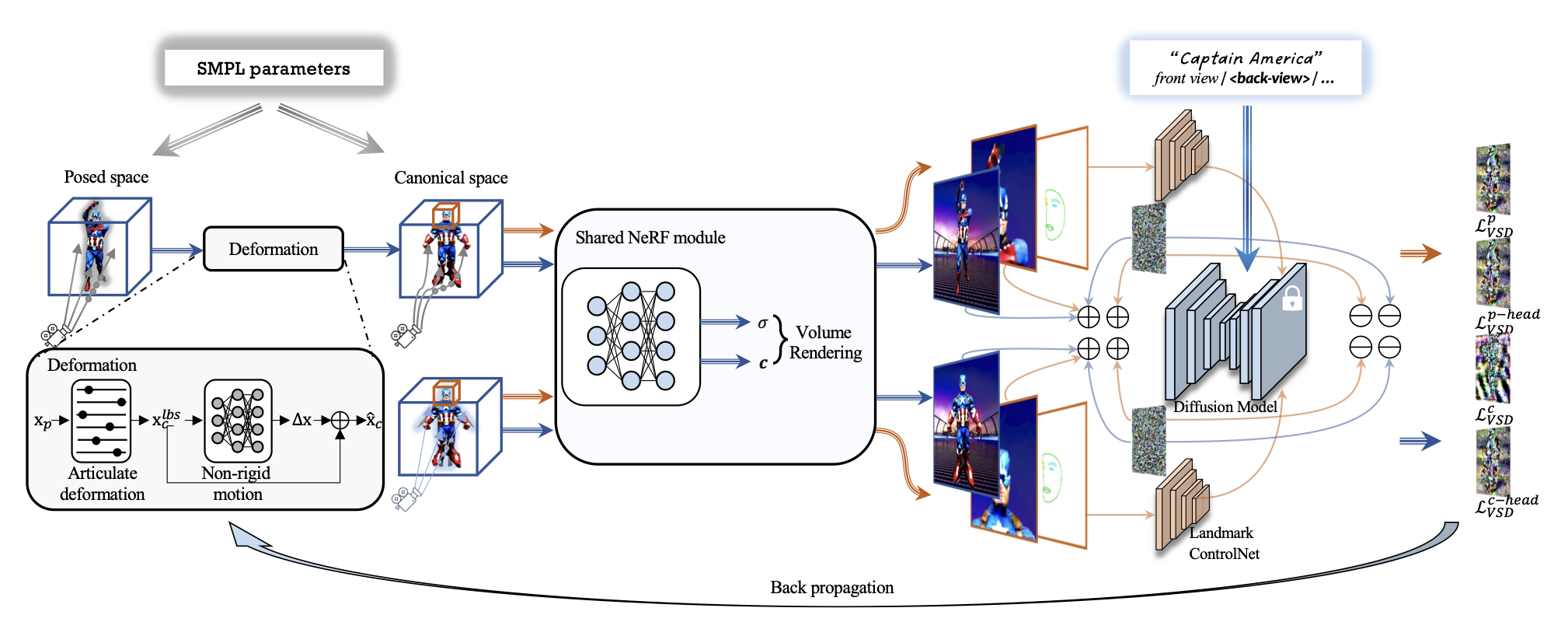}

	\caption{Overview of DreamAvatar.}
	\label{fig:03}
\end{figure}

\begin{figure}[t]
	\centering

	\includegraphics[width=\linewidth]{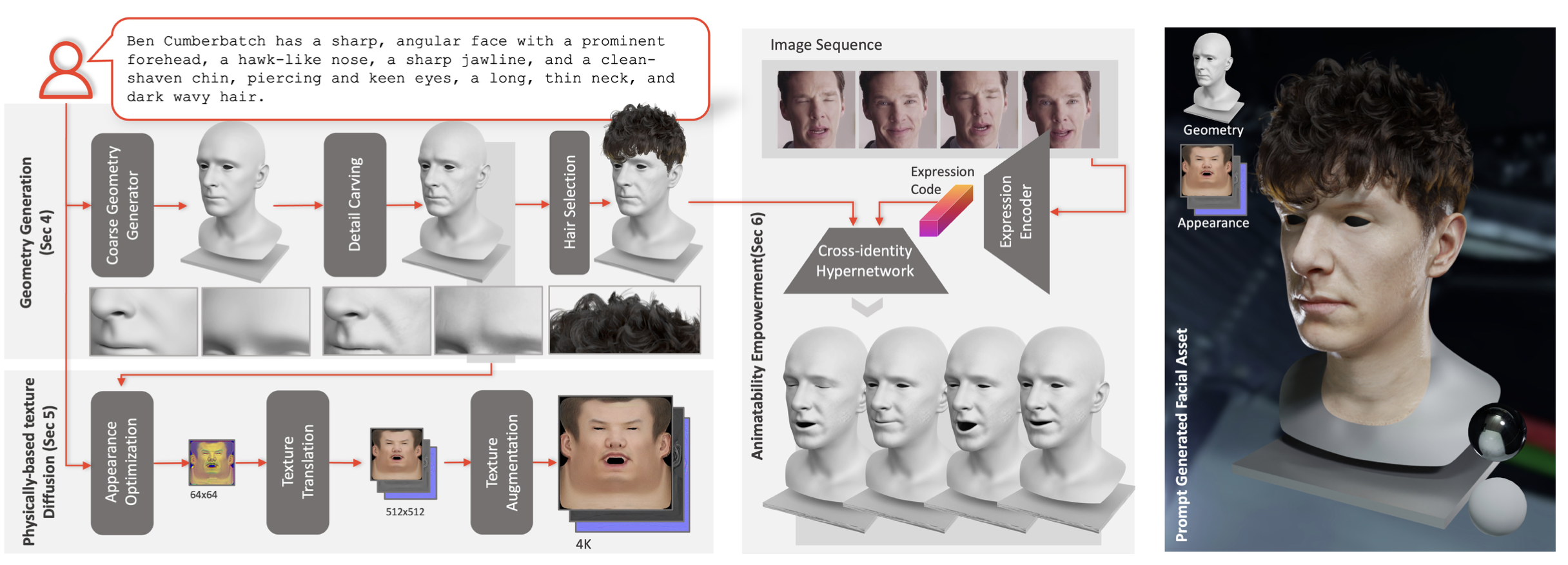}

	\caption{We show the pipeline of Dreamface~\cite{zhang2023dreamface}.}
	\label{fig:04}
\end{figure}

\begin{figure}[t]
	\centering

	\includegraphics[width=\linewidth]{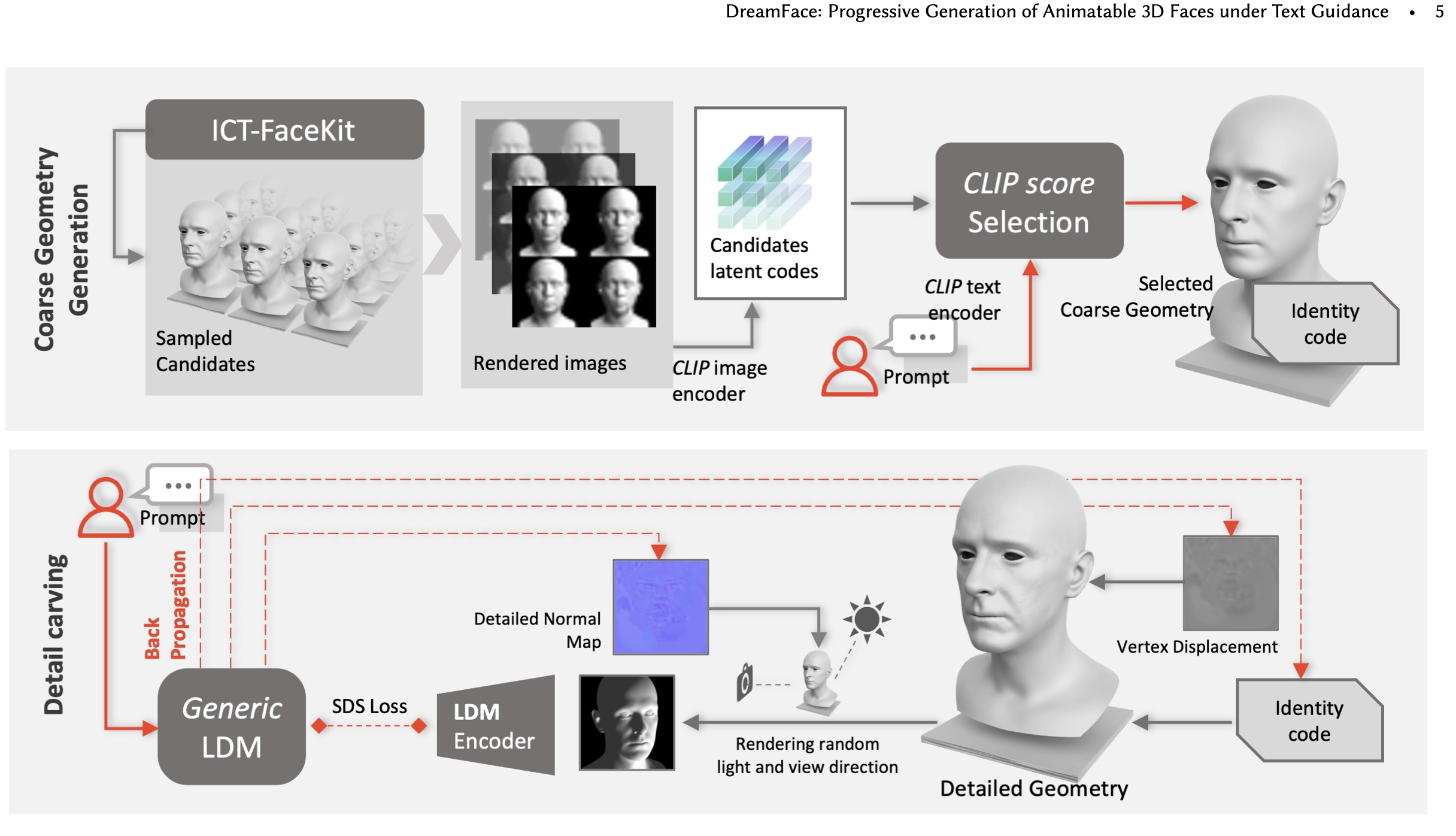}

	\caption{Geometry generation pipeline of DreamFace.}
	\label{fig:05}
\end{figure}

\begin{figure}[t]
	\centering

	\includegraphics[width=\linewidth]{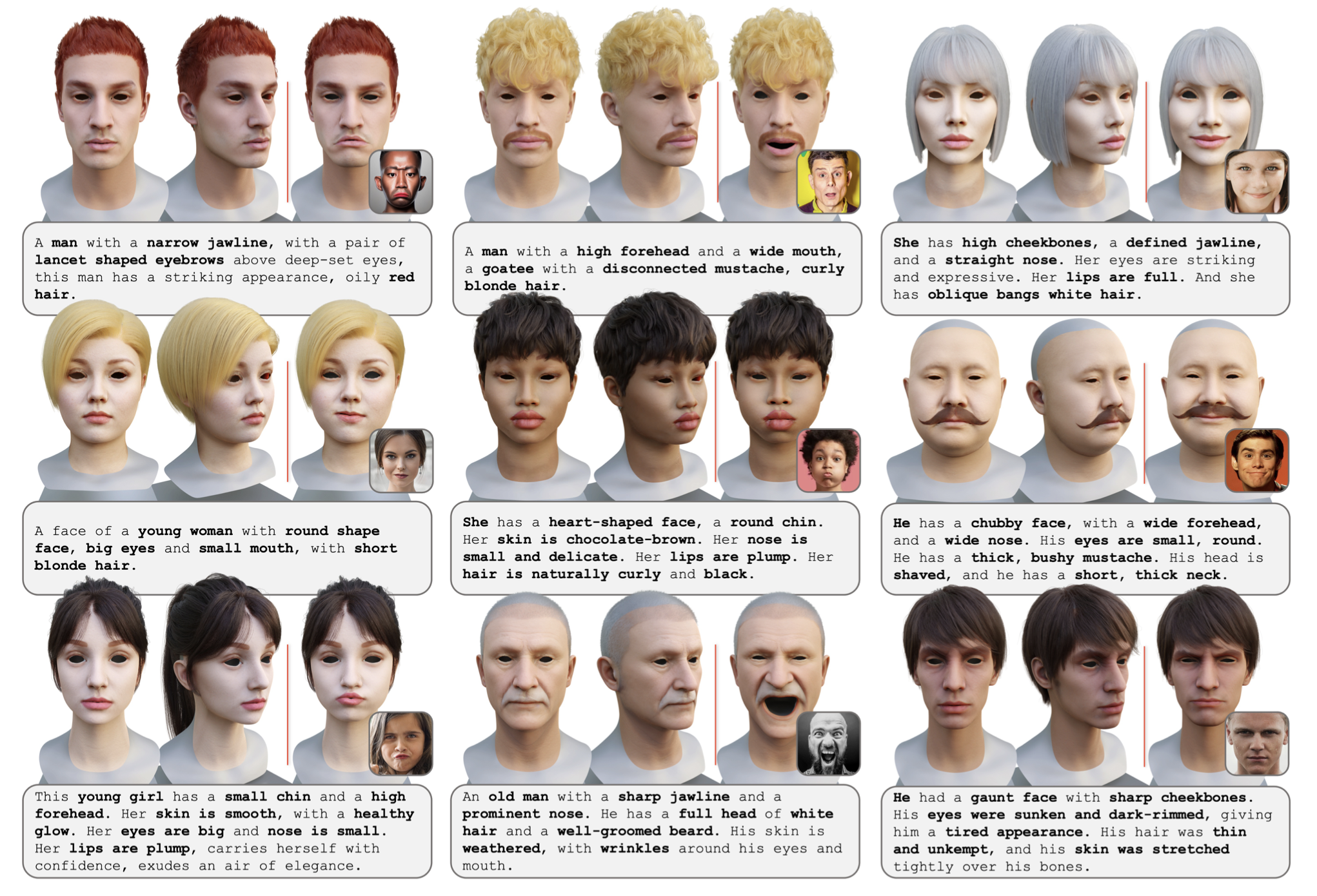}

	\caption{DreamFace generates facial assets of celebrities that capture their personalized characteristics and achieve a high degree of resemblance.}
	\label{fig:06}
\end{figure}

\begin{figure}[t]
	\centering

	\includegraphics[width=\linewidth]{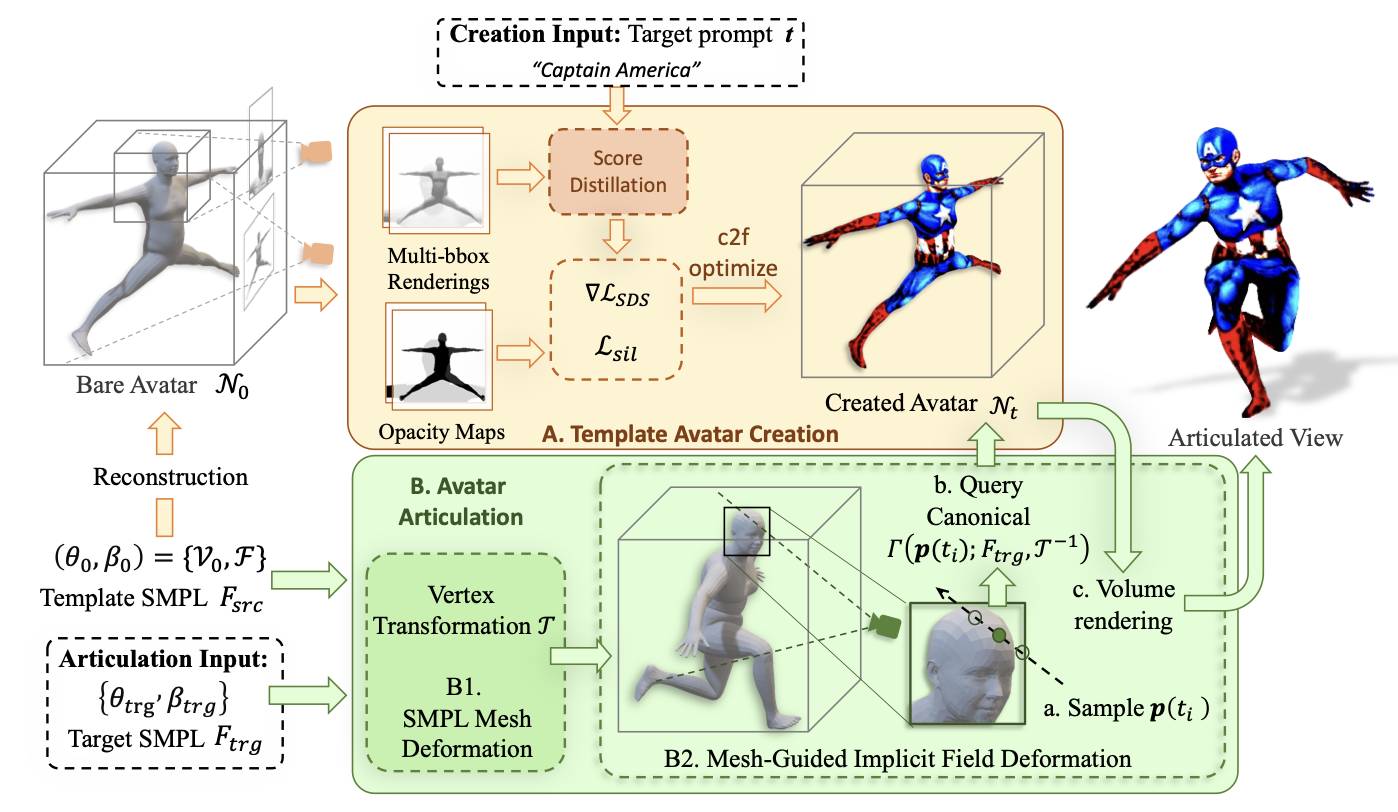}

	\caption{Method Overview of AvatarCraft~\cite{jiang2023avatarcraft}.}
	\label{fig:07}
\end{figure}

\begin{figure}[t]
	\centering

	\includegraphics[width=\linewidth]{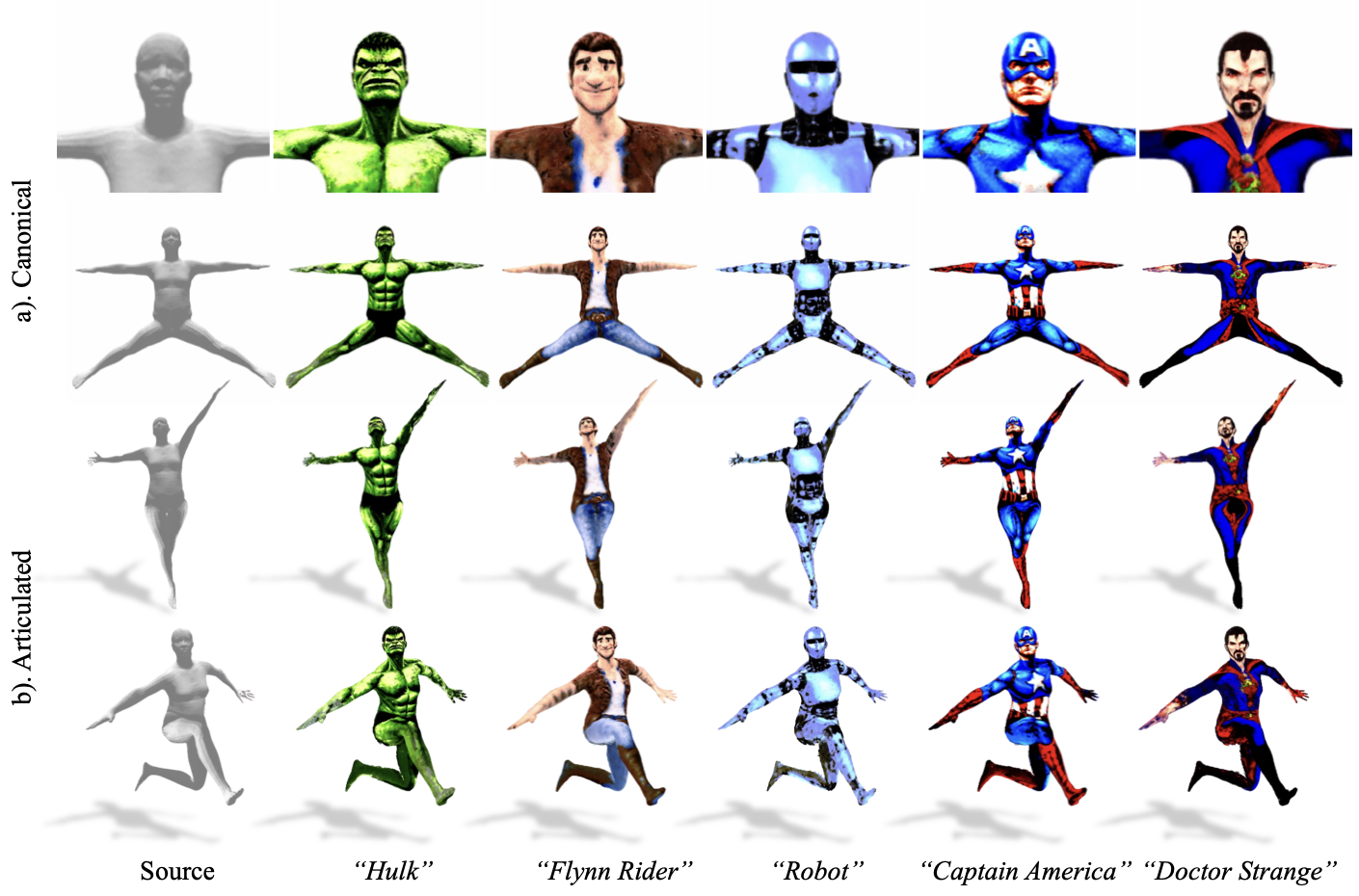}

	\caption{Qualitative Result of AvatarCraft.}
	\label{fig:08}
\end{figure}

\section{Avatar}

Text-to-3D~\cite{foo2023ai,li2024art3d,poole2022dreamfusion,lin2023magic3d,li2024generating,chan2023generative,tang2023make,shen2021deep,wang2023prolificdreamer,li2023sweetdreamer,metzer2023latent} is an emerging interdisciplinary field that focuses on converting textual descriptions into three-dimensional models. This technology has a wide range of applications, including virtual reality, gaming, and design. 

In recent years, the creation of 3D graphical human models has drawn considerable attention due to its extensive applications in areas such as movie production, video gaming, AR/VR and human-computer interactions, and the creation of 3D avatars through natural language could save resources and holds great research prospects. 3D human generation is a complex and evolving field in computer graphics and artificial intelligence, aiming to create detailed, animatable human models with realistic movements and non-rigid components such as hair and clothing. The work in this area can be broadly divided into two categories: generating the full 3D human body and generating the 3D avatar head. Below, we discuss significant contributions and methodologies in these domains.

The generation of 3D human avatars involves creating diverse virtual humans with different identities, shapes, and poses. This task is particularly challenging due to the variety in clothed body shapes and their complex articulations. Additionally, 3D avatars should be animatable, making the task even more difficult as it requires creating non-rigid components.

Early work in 3D human mesh generation~\cite{anguelov2005scape,joo2018total,osman2020star} utilized human parametric models. These models represent human body shapes using a small set of parameters that deform a template mesh. While these models facilitate easy synthesis of human shapes by adjusting a few parameters, they often fail to capture finer details due to the fixed mesh topology and limited resolution. 

To address these limitations, more recent methods have employed implicit non-rigid representations. For instance, SNARF~\cite{chen2021snarf} generate detailed 3D human avatars by predicting skinning fields and surface normals. gDNA~\cite{chen2022gdna} extends this approach by predicting a skinning field and normal field using an MLP, trained on raw posed 3D scans. AvatarGen~\cite{zhang2022avatargen} leverages an SDF based on a SMPL prior to provide disentangled control over geometry and appearance, while GNARF~\cite{bergman2022generative} and utilize generative and compositional NeRFs for high-resolution rendering and efficient training. Another approach focuses on creating animatable 3D avatars from 3D scans. Traditional methods often used linear blend skinning, which is simple but cannot produce pose-dependent variations, leading to artifacts. To overcome this, techniques like dual quaternion blend skinning~\cite{kavan2008geometric} and multi-weight enveloping~\cite{wang2002multi} have been developed.

Recently, implicit representations have become popular for generating animatable 3D avatars from scans due to their resolution-independence and smoothness. NASA~\cite{deng2020nasa} uses an occupancy field to model humans as deformable components.
Several method~\cite{chen2021snarf,tiwari2021neural} explore occupancy and SDF-based representations for animatable human characters.

Further advancements include deformable and animatable NeRFs, such as those proposed by previous works~\cite{peng2021animatable,liu2021neural,zheng2022structured}, which synthesize humans from novel poses and views.

Generating 3D avatar heads requires fine-grained control over facial expressions and realistic modeling of complex components like hair. Traditional methods like the 3D Morphable Models~\cite{blanz2023morphable} simplify face modeling by fitting values in a linear subspace. Variants include multilinear models~\cite{vlasic2006face}, and fully articulated head models~\cite{paysan20093d}. Deep generative methods~\cite{tran2019learning} and recent work has also focused on editing avatar faces. For example, IDE-3D~\cite{sun2022ide} allows interactive editing of shape and texture.

The generation of 3D human bodies and heads involves a variety of techniques ranging from parametric models and implicit representations to advanced neural network-based methods. These approaches strive to create detailed, realistic, and animatable 3D human models, addressing challenges like capturing fine details, managing non-rigid components, and ensuring realistic motion generation.

\section{Text-to-3D avatar generation}
\subsection{DreamAvatar}
DreamAvatar~\cite{cao2023dreamavatar} is a cutting-edge 3D human avatar generation framework. As shown in Fig, 1, 2 and 3, this framework employs a trainable Neural Radiance Field (NeRF) model to predict density and color for 3D points, complemented by pretrained text-to-image diffusion models that offer 2D self-supervision. The key innovations of DreamAvatar are threefold:

\paragraph{Utilization of the SMPL Model:} By leveraging the Skinned Multi-Person Linear (SMPL) model, DreamAvatar provides robust shape and pose guidance, facilitating the generation process to produce 3D models that adhere to human body structures.

\paragraph{Dual-Observation Space Design:} The framework introduces a novel design that incorporates two observation spaces—a canonical space and a posed space—related by a learnable deformation field, aiding in the generation of more comprehensive textures and geometries that are faithful to the target pose.

\paragraph{Joint Optimization of Loss Functions:} DreamAvatar optimizes not only from the perspective of the full body but also from a zoomed-in 3D head viewpoint, addressing the multi-face "Janus" problem and enhancing facial details in the generated avatars.

DreamAvatar details the implementation of the method, including the use of threestudio for NeRF and Variational Score Distillation (VSD), as well as specific versions of Stable Diffusion and ControlNet. Extensive evaluations on various text prompts have demonstrated DreamAvatar's effectiveness, outperforming existing text-guided 3D generation methods by producing 3D human avatars with high-quality geometry and consistent textures. Additionally, the paper presents user studies and further analyses that substantiate DreamAvatar's leading position in 3D avatar generation. It also discusses the limitations of the current implementation, such as the lack of consideration for animation and potential biases inherited from the pretrained diffusion model. Finally, the paper contemplates the societal impact of this technology, highlighting both the positive contributions to metaverse development and the risks associated with the malicious use of such technology.

\subsection{DreamFace}
DreamFace~\cite{zhang2023dreamface}, a pioneering framework introduced, represents a significant stride in the generation of 3D facial assets. As shown in Fig. 4, 5 and 6, this technology stands out for its ability to transform textual prompts into lifelike and customizable 3D faces with intricate details and animations. The methodology underpinning DreamFace is a testament to its innovative approach, combining a coarse-to-fine geometry generation strategy with a dual-path mechanism that employs both a generic Latent Diffusion Model (LDM) and a specialized texture LDM. This dual-path mechanism is pivotal as it ensures the generation of facial textures that are not only diverse but also adhere to the specific requirements of the UV space.

The framework's two-stage optimization process is another keystone of its success, allowing for efficient and fine-grained synthesis. This process is carried out in both the latent and image spaces, which is crucial for mapping the compact latent space to physically-based textures. The result is a seamless integration of high-quality textures that are essential for photo-realistic rendering.

Moreover, DreamFace's animation capabilities are enhanced through the support of blendshapes and the introduction of a cross-identity hypernetwork. This empowers the generation of personalized and nuanced facial expressions, enabling the creation of dynamic and engaging characters. The framework's compatibility with existing computer graphics pipelines and its ability to animate using in-the-wild video footage significantly democratize the creation of 3D facial assets, making it accessible to both professionals and novices alike.

DreamFace's methodological sophistication lies in its multi-stage generation process that progressively refines the facial geometry, meticulously applies physically-based textures, and enriches the asset with animatable features. This comprehensive approach positions DreamFace at the forefront of 3D facial asset generation technology, offering a versatile solution for a myriad of applications across the digital content creation spectrum.

\subsection{AvatarCraft}
As shown in Fig. 7 and 8, AvatarCraft~\cite{jiang2023avatarcraft} is an innovative approach to transforming text descriptions into high-quality 3D human avatars with controllable poses and shapes. This method leverages diffusion models to guide the learning of geometry and texture for a neural avatar based on a single text prompt.

AvatarCraft begins with a basic neural human avatar template. Given a text prompt, It uses diffusion models to update the template with geometry and texture that align with the text description. To address the challenge of distorted geometry and textures that can arise from direct application of diffusion models, the team designed a novel coarse-to-fine multi-bounding-box training strategy. This strategy improves global style consistency while preserving fine details.

A key aspect of AvatarCraft is shape regularization, introduced to stabilize the optimization process by penalizing the accumulated ray opacity of the avatar. This approach helps to prevent the generation of degenerate solutions such as empty volumes or flat geometry.

For animation capabilities, AvatarCraft employs a deformation technique that uses an explicit warping field. This field maps the target human mesh to a template human mesh, with both meshes represented using parametric human models. This method simplifies the animation and reshaping of the generated avatar by controlling pose and shape parameters, without the need for additional training.

The AvatarCraft framework also introduces a coarse-to-fine generation strategy that captures style details at different scales. This strategy is critical for generating avatars with important visual features correctly aligned, as it ensures that finer texture details are generated as the resolution increases.

In terms of implementation, the team combined the neural implicit field architecture of NeuS with the fast training approach of Instant-NGP. This combination improves reconstruction speed while maintaining quality. The avatar creation process involves training on multi-view renderings of a bare SMPL mesh, using the Score Distillation Score (SDS) loss from DreamFusion for guidance.

Extensive experiments using various text descriptions demonstrate that AvatarCraft is effective and robust in creating human avatars and rendering novel views, poses, and shapes. The project's webpage provides a more immersive view into the 3D results generated by the method.

\section{Applications}

The development of advanced techniques for 3D human and avatar generation has opened up a myriad of applications across various industries. These applications leverage the ability to create detailed, animatable, and realistic virtual humans for purposes ranging from entertainment to healthcare. Below, we explore several key areas where 3D human and avatar technologies are making a significant impact.

\subsection{Entertainment and Media}

One of the most prominent applications of 3D human and avatar generation is in the entertainment and media industry. Virtual humans are used extensively in video games, movies, and virtual reality (VR) experiences. In video games, 3D avatars provide players with customizable characters that enhance immersion and personal engagement. Games like "The Sims" and "Cyberpunk 2077" showcase highly detailed and customizable avatars, contributing to a richer player experience.

In the film industry, 3D avatars allow for the creation of realistic digital doubles of actors. These digital doubles can perform stunts, de-age actors, or even resurrect deceased performers, as seen with characters like Princess Leia in "Rogue One: A Star Wars Story." Moreover, fully animated movies like "Avatar" and "Toy Story" rely heavily on 3D character generation to bring fantastical worlds and characters to life.

Virtual reality and augmented reality (AR) also benefit from 3D avatars. In VR, users can interact with fully immersive environments where avatars represent themselves and others, enhancing social presence and interaction. In AR, applications like Snapchat filters and virtual try-on for fashion use 3D avatars to overlay virtual objects onto the real world.

\subsection{Social Media and Communication}

Social media platforms and communication tools are increasingly incorporating 3D avatars to enhance user interaction. Platforms like Facebook and Snapchat allow users to create personalized avatars, which can be used in posts, stories, and as virtual representations in online meetings. These avatars add a layer of personalization and expression, enabling users to convey emotions and personalities more effectively than through text or 2D images alone.

In communication, 3D avatars are used in virtual meetings and telepresence systems. Companies like Spatial and Zoom are developing technologies that allow users to join meetings as 3D avatars, creating a more engaging and interactive experience compared to traditional video calls. These avatars can mimic the user’s gestures and facial expressions, making remote communication feel more natural and connected.

\subsection{E-commerce and Retail}

In the e-commerce and retail sectors, 3D avatars are revolutionizing the way consumers shop online. Virtual try-on solutions allow customers to see how clothes, accessories, and even cosmetics look on a 3D model of themselves before making a purchase. This technology improves the shopping experience by helping consumers make more informed decisions and reducing the rate of returns due to sizing and fit issues.

For example, companies like Warby Parker and Sephora use augmented reality to let users try on glasses and makeup virtually. Similarly, fashion retailers are developing virtual fitting rooms where customers can upload a 3D scan of their body to try on clothing. This not only enhances the shopping experience but also allows for more personalized recommendations and a deeper connection between brands and consumers.

\subsection{Healthcare and Rehabilitation}

The healthcare industry is leveraging 3D human and avatar technologies for various applications, including medical training, patient care, and rehabilitation. In medical training, realistic 3D avatars can simulate complex surgical procedures, allowing students and professionals to practice in a controlled, risk-free environment. These simulations can include detailed anatomy and realistic physiological responses, providing a valuable learning tool.

For patient care, 3D avatars are used in telemedicine to facilitate remote consultations. Doctors can use avatars to explain medical conditions and treatments in a more visual and understandable way. In rehabilitation, virtual avatars are used in therapeutic exercises, helping patients recover from injuries or surgeries. For instance, patients can perform physical therapy exercises guided by a virtual avatar that demonstrates the correct movements and provides feedback.

\subsection{Fitness and Wellness}

3D avatars are also finding applications in the fitness and wellness industry. Virtual fitness trainers use avatars to guide users through workout routines, providing a more engaging and personalized experience. These trainers can adapt workouts based on the user’s performance and goals, offering real-time feedback and motivation.

Moreover, wellness apps are incorporating avatars to help users visualize their progress in areas like weight loss or muscle gain. By creating a 3D model that evolves with the user’s physical changes, these apps provide a tangible representation of progress, which can be highly motivating.

\subsection{Education and Training}

In the field of education and training, 3D avatars are used to create immersive learning environments. Virtual classrooms and training simulations can incorporate avatars to represent students and instructors, facilitating interactive and collaborative learning experiences. For example, language learning apps can use avatars to create conversational scenarios where learners practice speaking with virtual characters.

Corporate training programs also use 3D avatars to simulate real-world scenarios, such as customer service interactions or emergency response drills. These simulations provide employees with hands-on experience in a safe and controlled setting, improving their skills and confidence.

\subsection{Fashion and Personalization}

The fashion industry is leveraging 3D avatar technology to streamline design processes and enhance personalization. Designers use 3D avatars to create virtual prototypes of clothing, allowing for rapid iteration and visualization of designs before physical production. This not only speeds up the design process but also reduces waste by minimizing the need for physical samples.

For consumers, 3D avatars enable a new level of personalization. Custom clothing brands can use avatars to create perfectly fitted garments based on a 3D scan of the customer’s body. This technology ensures a better fit and more satisfaction with the final product.

\subsection{Cultural Heritage and Preservation}

In the field of cultural heritage, 3D avatars are used to recreate historical figures and scenes. Museums and educational institutions use these avatars to bring history to life, providing visitors with interactive and engaging experiences. Virtual tours can feature avatars of historical figures who guide visitors through exhibits, sharing stories and insights.

Moreover, 3D avatars are used in the preservation of cultural heritage sites. By creating detailed 3D models of sites and artifacts, researchers can study and preserve these cultural treasures for future generations. These models can be used for virtual reconstructions, allowing people to explore historical sites in their original form.

The applications of 3D human and avatar generation are vast and diverse, impacting numerous industries and aspects of daily life. From enhancing entertainment and social media experiences to revolutionizing e-commerce, healthcare, and education, the ability to create realistic and animatable 3D humans offers significant benefits. As technology continues to advance, we can expect even more innovative uses of 3D avatars, further integrating them into our digital and physical worlds.

\section{Conclusion}
In conclusion, the integration of AI with 3D human model and avatar generation has revolutionized digital content creation, offering higher quality and broader access. The paper emphasizes the technology of personalized avatar creation and the significant applications across industries. While the technology promises significant benefits, it also presents challenges that need to be managed responsibly. As these advancements continue, they are set to reshape our digital experiences and interactions.

\bibliographystyle{citation}
\bibliography{citation}

\end{document}